\documentclass[%
superscriptaddress,
reprint,
 amsmath,amssymb,
 aps,
]{revtex4-2}

\usepackage{color}

\newcommand{\bibcolor}[1]{\textcolor{black}{#1}}

\newcommand{\markupcolor}{black}

\usepackage{graphicx, placeins, float}
\usepackage{dcolumn}
\usepackage{bm}
\usepackage{physics}

\usepackage{hyperref}

\usepackage{amsmath}

\usepackage{amsmath}
\newcommand\numberthis{\refstepcounter{equation}\tag{\theequation}}

\begin{document}

\title{End-to-end variational quantum sensing}

\author{Benjamin MacLellan}
\email{benjamin.maclellan@uwaterloo.ca}
\affiliation{University of Waterloo, Department of Physics \& Astronomy, 200 University Ave., Waterloo, ON, Canada}
\affiliation{Institute for Quantum Computing, 200 University Ave., Waterloo, ON, Canada}
\affiliation{Perimeter Institute for Theoretical Physics, 31 Caroline St N., Waterloo, ON, Canada}
\affiliation{Ki3 Photonics Technologies, 2547 Rue Sicard, Montreal, QC, Canada}

\author{Piotr Roztocki}
\affiliation{Ki3 Photonics Technologies, 2547 Rue Sicard, Montreal, QC, Canada}

\author{Stefanie Czischek}
\affiliation{University of Ottawa, Department of Physics, 75 Laurier Ave E, Ottawa, ON, Canada}

\author{Roger G. Melko}
\affiliation{University of Waterloo, Department of Physics \& Astronomy, 200 University Ave., Waterloo, ON, Canada}
\affiliation{Perimeter Institute for Theoretical Physics, 31 Caroline St N., Waterloo, ON, Canada}

\date{\today}

\begin{abstract}
Harnessing quantum correlations can enable sensing beyond classical precision limits, with the realization of such sensors poised for transformative impacts across science and engineering.
Real devices, however, face the accumulated impacts of noise and architecture constraints, making the design and success of practical quantum sensors challenging.
Numerical and theoretical frameworks to optimize and analyze sensing protocols in their entirety are thus crucial for translating quantum advantage into widespread practice.
Here, we present an end-to-end variational framework for quantum sensing protocols, where parameterized quantum circuits and neural networks form trainable, adaptive models for quantum sensor dynamics and estimation, respectively.
The framework is general and can be adapted towards arbitrary qubit architectures, as we demonstrate with experimentally-relevant ansätze for trapped-ion and photonic systems, and enables to directly quantify the impacts that noise and finite data sampling.
End-to-end variational approaches can thus underpin powerful design and analysis tools for practical quantum sensing advantage. 
\end{abstract}

\maketitle

\section{Introduction \label{sec:introduction}}

\begin{figure}[t]
    \centering
    \includegraphics[width=88mm]{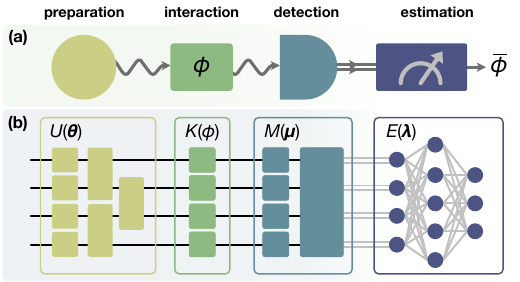}
    \caption{\textbf{Building a quantum sensor.} (a) The value of an unknown parameter $\phi$ is estimated from measurements on a probe state. Classical limits on sensing precision can be surpassed in certain operational regimes, pending the careful design and realization of all protocol steps. (b) Quantum sensing protocols lend themselves well to being framed as variational optimization problems, with end-to-end parameterized representations enabling powerful design and analysis tools. In this framing, the probe state is prepared via a parameterized unitary operation $U(\bm{\theta})$; it interacts with the parameter of interest $\phi$ through a quantum channel, $K(\phi)$; and is read-out to classical information by a parameterized measurement $M(\bm{\mu})$. A classical neural network, $E(\bm{\lambda})$, then estimates the parameter value, $\bar{\phi}$, from a finite set of measurement data.
    }
    \label{fig:concept}
\end{figure} 

By leveraging carefully-prepared quantum correlations, sensing protocols that surpass the intrinsic precision bounds of classical approaches can be designed \cite{giovannettiQuantumenhancedMeasurementsBeating2004}.
Such protocols comprise four steps (Fig. 1a): (i) the preparation of a metrologically-useful probe state; (ii) the interaction of the probe with an unknown physical parameter of interest $\phi$; (iii) the measurement of the perturbed probe state, producing a dataset of classical detection outcomes; and (iv) the estimation $\bar{\phi}$ of the physical parameter from the classical dataset. 
This approach underpins the many applications of quantum sensing, spanning gravitational wave detection \cite{schnabelQuantumMetrologyGravitational2010a}, earth science \cite{strayQuantumSensingGravity2022}, illumination \cite{lopaevaExperimentalRealizationQuantum2013}, microscopy \cite{taylorBiologicalMeasurementQuantum2013a}, energy engineering \cite{crawfordQuantumSensingEnergy2021}, and chemical dynamics \cite{ciminiQuantumSensingDynamical2019}, among others.

In recent years, approaches for the design and analysis of quantum sensing protocols have rapidly advanced  
\cite{rubioNonasymptoticQuantumMetrology2021,meyerQuantumMetrologyFinitesample2023, zhouAchievingHeisenbergLimit2018}.
Moreover, as variational and machine learning approaches become ubiquitous across the landscape of quantum technologies \cite{torlaiNeuralnetworkQuantumState2018, metzSelfcorrectingQuantumManybody2023, krennArtificialIntelligenceMachine2023, youssryExperimentalGrayboxQuantum2024, cerezoVariationalQuantumAlgorithms2021}, recent efforts have begun to explore their impact in the context of quantum sensing \cite{huangQuantumMetrologyAssisted}.
For example, variational quantum-classical algorithms have been utilized to drive programmable quantum devices into regimes suited for sensing tasks [\citenum{meyerVariationalToolboxQuantum2021, beckeyVariationalQuantumAlgorithm2022}, 
\textcolor{\markupcolor}{\citenum{yangVariationalPrincipleOptimal2022, koczorVariationalstateQuantumMetrology2020, kaubrueggerVariationalSpinSqueezingAlgorithms2019}}] 
\textcolor{\markupcolor}{and have been implemented on experimental architectures, e.g., atom arrays }\cite{marciniakOptimalMetrologyProgrammable2022}. 
\textcolor{\markupcolor}{
    Numerical techniques including
    tensor networks \cite{kaubrueggerVariationalSpinSqueezingAlgorithms2019, chabudaTensornetworkApproachQuantum2020,kurdzialekQuantumMetrologyUsing2024}, 
    and higher-order operations \cite{bavarescoDesigningOptimalProtocols2024}
    have been used to represent quantum states, dynamics, and measurements in the design of optimal quantum sensing protocols; 
    while learning and optimization techniques encompass 
    see-saw methods \cite{macieszczakBayesianQuantumFrequency2014, tothActivatingHiddenMetrological2020, trenyiActivationMetrologicallyUseful2024},
    semidefinite programming \cite{tothQuantumStatesPositive2018, kurdzialekQuantumMetrologyUsing2024},
    evolutionary algorithms \cite{hentschelEfficientAlgorithmOptimizing2011,koczorVariationalstateQuantumMetrology2020}, 
    global optimization \cite{kaubrueggerVariationalSpinSqueezingAlgorithms2019}, 
    gradient descent \cite{meyerVariationalToolboxQuantum2021,beckeyVariationalQuantumAlgorithm2022}, 
    and reinforcement learning \cite{belliardoModelawareReinforcementLearning2024}.
}
    Further, classical machine learning methods, trained on real or simulated data from quantum devices, have also been proposed for performing the estimation step of sensing protocols in a Bayesian context \cite{greplovaQuantumParameterEstimation2017, nolanMachineLearningApproach2021}.
\textcolor{\markupcolor}{
    Finally, adaptive quantum sensing protocols -- in which subsequent measurement settings are conditioned on previous outcomes -- have also been studied in the context of learning and optimization methods, including with evolutionary algorithms \cite{hentschelEfficientAlgorithmOptimizing2011}, tensor networks \cite{kurdzialekQuantumMetrologyUsing2024},
    and reinforcement learning \cite{belliardoModelawareReinforcementLearning2024}.
}

Such approaches have helped identify the operational regimes where quantum sensing advantage may be found, but challenges remain in translating such insights into useful, near-term technologies.
Noise sources, fabrication tolerances, constraints on quantum control (e.g., qubit connectivity, native operations), finite sampling rates, as well as unequal interaction of $\phi$ across all parts of the probe state all arise in practical realizations of quantum sensing.
\textcolor{\markupcolor}{Thus far, studies have largely focused on the impact of such effects on one or more protocol stages at a time by, e.g., quantifying the decrease in state preparation fidelity or quantum Fisher Information in the presence of noise or imperfect operations \cite{dattaQuantumMetrologyImperfect2011}.}
In practice, deviations from ideal behaviour affect all protocol stages from preparation through to estimation, impacting downstream stages of the protocol in non-trivial ways due to the tight interdependence of operational parameters. 
\textcolor{\markupcolor}{Importantly, if the impacts of experimental constraints and imperfections are not quantified in terms of final estimator performance, effective mitigation strategies that seek to maintain quantum advantage cannot be developed.}
This makes `end-to-end' approaches promising for the design and study of practical quantum sensing systems, i.e., numerical and theoretical frameworks that support simulation and optimization across all protocol stages, with direct and quantifiable impacts on estimator performance.

Here, we introduce a numerical workflow that unifies preparation, interaction, measurement, and estimator parameterization under one end-to-end variational framework (Fig. 1b).
In Section \ref{subsec:device}, we examine how quantum sensor dynamics can be modelled as parameterized quantum circuits. By computing and optimizing the Fisher Information as a function of circuit parameters, we converge the sensor towards regimes that support maximal achievable estimation precision.
In Section \ref{subsec:bayes}, we present the training of a classical neural network as a variational estimator of the unknown parameter $\phi$ and, using a Bayesian approach to update the estimator as more data is observed, efficiently utilize the finite data sampling rates of quantum devices
In Section \ref{subsec:ansatz}, we demonstrate the full end-to-end variational quantum sensing (VQS) protocol on probe state ansätze relevant to trapped-ion and photonic platforms, and benchmark the Fisher Information, estimator bias, and estimator variance of the learned protocols.
Finally, in Section \ref{subsec:scaling} we benchmark the scaling performance of the full framework, and close with Section \ref{subsec:ghz} where we compare the effects of noisy probe state preparation for the GHZ and VQS protocols.

We find that the VQS approach enables the identification of optimal probe state and measurement regimes in the presence of realistic experimental constraints, as well as the direct study of how protocol imperfections and finite data effects impact estimation.
These functionalities empower the development of strategies for realizing and maintaining quantum sensing advantage in practical, near-term devices.

\section{Results \label{sec:discussion}}

\begin{figure*}[t]
    \centering 
    \includegraphics[width=180mm]{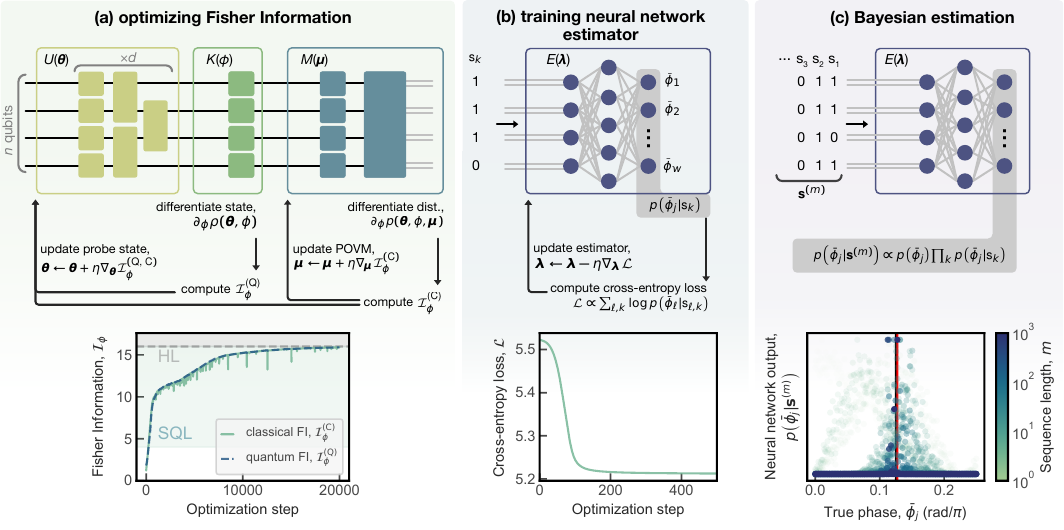}
    \label{fig:vqs_flow}
    \caption{\textbf{End-to-end optimization and training of variational quantum sensing protocols.} (a, top) The quantum sensor dynamics are modeled as a parameterized circuit, where the probe state preparation and detection parameters ($\bm{\theta}$ and $\bm{\mu}$, respectively) are tuned to optimize the quantum and classical Fisher Information with respect to $\phi$. (a, bottom) An example optimization curve using the CFI as the objective function and tracking the QFI throughout, which both converge to the HL. (b, top) A dense, feed-forward neural network is used as a variational ansatz for the estimator. The output layer of the network is interpreted as the posterior probability distribution of the phase value, $\bar{\phi}$, conditioned on measurement outcome $s_i$ \cite{nolanMachineLearningApproach2021} The network is trained on measurement data from the quantum device, sampled across $\phi$, in a (b, bottom) supervised classification mode using the cross-entropy loss function. (c, top) Using a Bayesian update rule, the network is used to estimate $\bar{\phi}$ from a sequence of measurements. This enables efficient training on individual samples and estimation on sequences of arbitrary length, $m$. (c, bottom) As the more data is observed, the neural network posterior distribution converges around the the true phase value, $\phi$ (black vertical line), reducing the bias of the estimated phase value $\bar{\phi}$ (\markupcolor vertical line).
    }
\end{figure*}




\subsection{Variational optimization of quantum sensor dynamics \label{subsec:device}}
Parameterized quantum circuits are a useful model for studying and designing quantum systems in a unified mathematical framework. Applications of such circuits include variational quantum algorithms,  learning ground states of many-body systems, among others \cite{cerezoVariationalQuantumAlgorithms2021, kandalaHardwareefficientVariationalQuantum2017, benedettiParameterizedQuantumCircuits2019}. 
For a quantum sensing task, the circuit representation of the sensor is divided into three parts. First, a parameterized unitary operation, $U(\bm{\theta})$, prepares a metrologically-useful probe state. This probe state interacts dynamically with the parameter of interest, $\phi$, through a quantum channel, $K(\phi)$. Throughout this work, we consider the task of single-parameter phase estimation, such that $K(\phi)$ is a local unitary rotation and $\phi$ is a single scalar value. However, this method can be generalized to multiparameter estimation, where $\phi$ is a real-valued vector, or to sensing tasks in which $K$ represents an alternative dynamical process, e.g., a non-unitary channel. After the probe state interacts with and is perturbed by the parameter $\phi$, the probe state is measured with a parameterized positive operator-valued measurement (POVM),
\(
    M(\bm{\mu}) = \{ \Pi_i(\bm{\mu}) | \sum_i \Pi_i(\bm{\mu}) = 1  \}.
\)
Here, we consider the elements of the POVM to be projective measurements in locally-rotated bases, which is realized via a layer of local rotation gates parameterized by $\bm{\mu}$, followed by measuring in the computational basis. Such measurements are, in general, easily accessible and can be performed with high fidelity in many qubit architectures. This POVM maps the density operator to a likelihood probability distribution over measurement outcomes, 
\[
    p( s_i | \phi ) = \text{Tr}\left[ \rho(\bm{\theta}, \phi) \Pi_i(\bm{\mu})  \right],
    \numberthis
\]
where $s_i$ is the $n$-bit measurement outcome associated with the projector, $\Pi_i$.
For a sensing protocol to realize the maximum precision that is possible, the state preparation and detection stages must be designed carefully around how the parameter-of-interest, $\phi$, interacts with the probe state. 

In the variational quantum sensing framework, the device parameters, $\bm{\theta}$ and $\bm{\mu}$, are tuned towards a regime where the probe state and its associated measurement can enable high-precision estimates of the value of $\phi$. Central to such a method is the choice of the objective function. For sensing tasks, the Fisher Information is a ubiquitous metric -- often used for quantifying the performance of a sensing apparatus \textcolor{\markupcolor}{by providing a lower bound on the achievable variance of any unbiased estimator}. The Fisher Information  quantifies the amount of information that a probability distribution or quantum state carries about a random variable which parameterizes it. These quantities are referred to as the classical (CFI) or quantum (QFI) Fisher Information, respectively, and both are closely related to the susceptibility of the distribution or state to infinitesimal perturbations, $\delta \phi$. 

The CFI of a discrete probability distribution $p(s_i | \phi)$ that is parameterized by $\phi$ and describes the probability of measurement outcomes, $\{ s_i \}$, may be expressed as,
\[
	\mathcal{I}_\phi^\text{(C)} = \sum_i \frac{(\partial_\phi p(s_i | \phi))^2 }{p(s_i | \phi)}
    . \numberthis
\]
Here, $\partial_\phi$ denotes the first-order partial derivative with respect to the parameter of interest, i.e. $\partial/\partial \phi$. The CFI can be viewed as the curvature (i.e., second-order derivative) of the Kullbeck-Leibler divergence between $p(\phi)$ and $p(\phi + \delta \phi)$. 

The QFI for an arbitrary, $n$-qubit quantum state, may be expressed as \cite{liuQuantumFisherInformation2019},
\[
	\mathcal{I}_\phi^{\text{(Q)}} 
    =
    \sum_{ \substack{i, i'=1 \\ \lambda_i + \lambda_{i'} \neq 0} }^{2^n}
    \frac{2 \text{Re}( \bra{\lambda_i} \partial_\phi \rho \ket{\lambda_{i'}} ) }{\lambda_i + \lambda_{i'}}
    , \numberthis
\]
where $\{ \lambda_i \}$ are the eigenvalues of the density operator.
The QFI is an important quantity in quantum information, with close relationships to the Bures metric \cite{safranekDiscontinuitiesQuantumFisher2017} and the quantum geometric tensor \cite{stokesQuantumNaturalGradient2020a}, as well as a useful metric for detecting entanglement in quantum systems \cite{haukeMeasuringMultipartiteEntanglement2016, hyllusFisherInformationMultiparticle2012, tothMultipartiteEntanglementHigh2012}. The QFI is equivalent to the optimal classical Fisher Information of a quantum state over all valid POVMs. As such, the probe state and measurement procedure must be tailored to each other and to the form of interaction with $\phi$.

For closed systems, where the probe state is pure and $K(\phi)$ is unitary, the QFI simplifies to \cite{liuQuantumFisherInformation2019},
\[
	\mathcal{I}_\phi^{\text{(Q)}} = 
    4 \text{Re} ( 
        \bra{\partial_\phi \rho}\ket{\partial_\phi \rho}
        - |\bra{\partial_\phi \rho}\ket{\rho}|^2
    )
    \ . \numberthis
\]

In the context of quantum sensing, the primary importance of the Fisher Information is that it lower bounds the variance of any unbiased estimator of the parameter $\phi$,
\[
    \Delta^2 \bar{\phi} \geq \frac{1}{\mathcal{I}_\phi}
    \ . \numberthis
\]
This inequality is the famous Cramér–Rao bound, and holds in both the quantum and classical contexts \cite{helstromMinimumMeansquaredError1967, tothQuantumMetrologyQuantum2014}.

Using only classical resources (i.e., no entanglement, squeezing, etc.), the Fisher Information can scale, at most, linearly in the size of the probe state -- e.g., for an $n$-qubit probe state $\mathcal{I}_\phi \sim n$. This scaling is known as the standard quantum limit (SQL) or shot-noise limit. However, if the probe state uses quantum resources such as entanglement, then the Fisher Information can scale with a quadratic improvement over the classical case, $\mathcal{I}_\phi \sim n^2$, which is known as the Heisenberg limit (HL).

As the first stage of the VQS protocol, we evaluate and optimize the Fisher Information of the parameterized quantum circuit (Fig. 2a). We use differentiable programming of the circuits to easily compute the quantum and classical Fisher Information with respect to $\phi$. First, a forward simulation pass of the circuit is performed to compute the perturbed quantum probe state, $\rho(\phi)$, and likelihood distribution $p(s_i | \phi)$. Using automatic differentiation, the derivative of the state and distribution with respect to the parameter of interest is computed, $\partial_\phi \rho$ and $\partial_\phi p(s_i | \phi)$, respectively, and used to compute the QFI and/or CFI. As differentiable programming allows the arbitrary stacking of derivative passes, the gradient of the FI with respect to $\bm{\theta}$ and $\bm{\mu}$ can be computed, i.e., 
$\nabla_{\bm{\theta}} \mathcal{I}_\phi^{\text{(Q)}}$, $\nabla_{\bm{\theta}} \mathcal{I}_\phi^{\text{(C)}}$, and/or $\nabla_{\bm{\mu}} \mathcal{I}_\phi^{\text{(C)}}$. By updating the values of the probe state and detection parameters via a gradient ascent optimization algorithm, the circuit can be tuned into a regime which has larger Fisher Information and, thus, maximum achievable estimation precision.

Once the circuit optimization has reached an appropriate stopping criterion (e.g., maximum number of iterations or plateau of the objective function), the parameters $\bm{\theta}$ and $\bm{\mu}$ are fixed. A dataset of measurement outcomes -- classical bitstrings, $s_i = \{ 0,1 \}^{n} $ -- are collected for training and benchmarking the estimator. 
A sequence of measurement outcomes of length $|D_\phi|$ are sampled at $n_\phi$ distinct phase values across the range of interest, $\phi \in [\phi_\text{min}, \phi_\text{max}]$. Thus, the total number of bitstrings in the training dataset is $|\mathcal{D}| = n_\phi \times |\mathcal{D}_\phi|$. This dataset can be viewed as a three-dimensional binary array of shape $(n_\phi, |\mathcal{D}_{\phi}|, n)$.

\begin{figure*}[tb]
    \centering 
    \includegraphics[width=180mm]{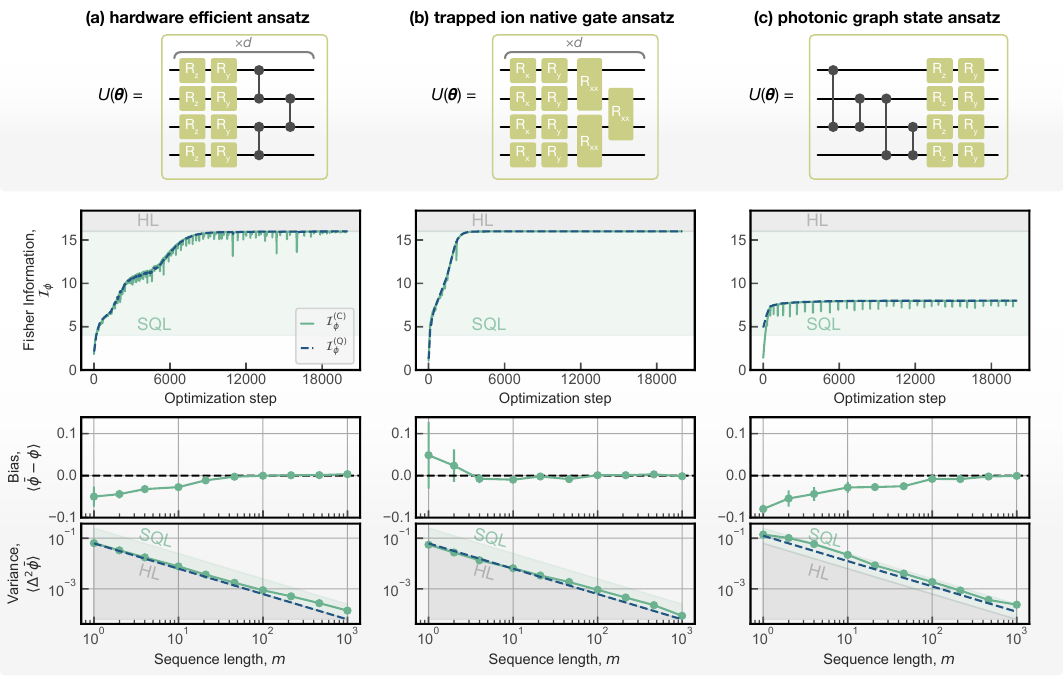} 
    \caption{\textbf{VQS with hardware-relevant ansätze.} We demonstrate the end-to-end framework with three unitary probe state preparation ansätze; including (a) a hardware-efficient ansatz (HEA) with controlled-phase interactions \cite{kandalaHardwareefficientVariationalQuantum2017},
    (b) trapped-ion native gate interactions (TIA) \cite{czischekSimulatingMeasurementinducedPhase2021}, 
    and (c) with states locally equivalent to a bundled graph state. 
    The Fisher Information of the HEA and TIA plateau to the HL \textcolor{\markupcolor}{for an $n$-qubit system}, the optimal value for entangled probe states, while the bundled graph state plateaus to a FI of $n^2/2$, the theoretically known QFI value for such states \textcolor{\markupcolor}{(with $n=4$)}. 
    The trained neural network estimators all converge towards being unbiased with increasing sequence length $m$, and partially saturate the CRB for the variance.  
    For all results presented the number of qubits is $n=4$ and the HEA and TIA circuit depth is $d=4$.
    }
    \label{fig:ansatze}
\end{figure*}

\subsection{Bayesian estimation via neural networks \label{subsec:bayes}}

With the quantum dynamics of the sensor optimized and datasets of measurement outcomes collected, we train a neural network ansatz as a variational estimation function. Here, we use the approach proposed by Nolan \textit{et al}. \cite{nolanMachineLearningApproach2021}, in which a dense, feed-forward neural network, $E(\bm{\lambda})$, outputs an estimate of the phase value, $\bar{\phi}$, given a sequence of measurement shots, $\bm{s}^{(m)}$. The network is composed of $n$ input nodes (corresponding to the projective measurement outcome on the $n$-th qubit) and $w$ output nodes, and is parameterized by weights and biases, $\bm{\lambda}$. The output layer of the network is normalized to be a valid probability distribution (via the softmax function) and interpreted as the posterior distribution, $p(\bar{\phi}_j | s_i)$, of estimating the phase value to be $\bar{\phi}_j$ given the single measurement outcome, $s_i$. 
We note that in contrast to previous works using neural network estimators \cite{nolanMachineLearningApproach2021}, we use the raw bitstrings sampled from the quantum device as the input to the network. This format allows the network to more easily capture the relevant correlations between qubits. We find that this leads to improved convergence and estimator performance.
As the network learns the correlations between $\phi$ and the measurement outcomes directly from data, it can better compensate for noisy probe state preparation in the estimation, as we explore later. 

The network is trained on single-shot measurements (encoded as one-hot vectors) via the cross-entropy loss function,
\[
	\mathcal{L} = \frac{1}{n_\phi |\mathcal{D}_\phi|} \sum_{\ell=1}^{n_\phi}\sum_{k=1}^{|\mathcal{D_\phi}|} \log p(\bar{\phi}_{\ell} | s_{k, \ell})
    . \numberthis
    \label{eqn:cross_entropy}
\]
Training the neural network can be viewed as indirectly performing Bayesian inversion, learning an approximate posterior distribution from sampled data. 
Once the network is trained, estimation is performed on sequences of arbitrary length shots via a Bayesian update rule. For a sequence of measurement outcomes of length $m$, $\bm{s}^{(m)} =  \{ s_1, s_2, \dots, s_m \}$, each individual shot is fed into the neural network and the normalized product of the output layers is interpreted as the posterior distribution,
\[
	p\left(\bar{\phi}_j | \bm{s}^{(m)} \right) \propto \prod_{k=1}^m p( \bar{\phi}_j | s_k )
    , \numberthis
    \label{eqn:bayesian_update}
\]
where we assume conditional independence of the samples and a uniform prior. The estimated value is,
\[
	\bar{\phi} = 
    \underset{\bar{\phi}_j}{\text{argmax}} \left[ p\left( \bar{\phi}_j | \bm{s}^{(m)} \right) \right],
    \label{eqn:estimated_phase}
    \numberthis
\]
and the bias and variance of this estimated value are computed as $\langle \bar{\phi} - \phi \rangle$ and
\[ 
    \Delta^2 \bar{\phi} = \sum_{j=1}^{w} p(\bar{\phi}_j | \bm{s}^{(m)}) \left( \bar{\phi} - \phi \right)^2
    , \numberthis
\]
respectively \cite{nolanMachineLearningApproach2021}.

In the design and benchmarking of metrology schemes, it is desired for estimators to be both unbiased and minimum variance; here, the Cramér–Rao bound provides an asymptotic lower bound of the variance for unbiased estimators. For an estimation procedures with repeated measurements, the Cramér–Rao bound is expressed as,
\(
    \Delta^2 \bar{\phi} \geq (m\mathcal{I}_\phi)^{-1}
\), 
such that $(m n)^{-1}$ and $(mn^2)^{-1}$ are the SQL and HL on estimating $\phi$ from $m$ measurements of an $n$-qubit system, respectively.
Statements about estimator performance are often framed asymptotically -- that is, in the assumption of an infinite number of measurements samples -- an assumption which often does not hold in real, experimentally-relevant contexts. Recent work has made significant progress in theoretical frameworks in limited-data regimes \cite{rubioNonasymptoticQuantumMetrology2021, meyerQuantumMetrologyFinitesample2023}; here we explicitly benchmark VQS protocols in the regime of finite data using synthetic data sampled from differentiable tensor network simulations of the quantum dynamics.

\begin{figure*}[tb]
    \centering 
    \includegraphics[width=180mm]{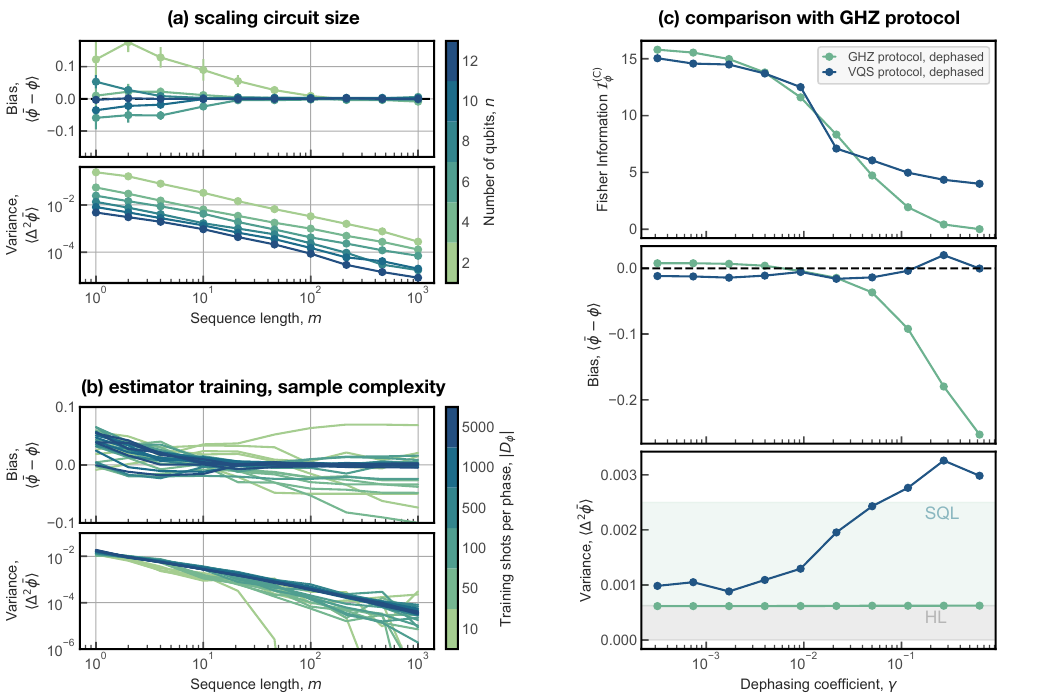} 
    \caption{\textbf{Scaling analysis of the VQS method.} (a) The VQS neural network estimator bias and variance for $n$-qubit probe state ansätze. (b) The required number of training shots per phase, $|\mathcal{D}_\phi|$, to robustly train the neural network estimator. (c) Comparison between GHZ and VQS protocols with dephasing noise in the state preparation, including the classical Fisher Information of the quantum device and the neural network estimator bias and variance. \textcolor{\markupcolor}{The GHZ protocol estimator becomes biased as noise in the probe state preparation increases, though the variance remains unchanged -- due to the analytical form of the maximum likelihood estimator (see Methods).} 
    On the other hand, the VQS method can tune the device parameters toward probe states that are more robust to the particular form of noise. The VQS neural network estimator remains unbiased as the noise increases, as the network can learn the effect of noise from the training data, and instead trades off for increased variance. For the results presented in (b) and (c), the number of qubits and circuit layers is $n=d=4$.
    }
    \label{fig:scaling}
\end{figure*}

\subsection{Hardware-relevant ansätze \label{subsec:ansatz}}
First, we demonstrate the VQS framework end-to-end with three ansätze that are relevant to current and near-term qubit architectures (Fig. 3). First, we consider a hardware-efficient ansatz (HEA) \cite{kandalaHardwareefficientVariationalQuantum2017}, composed of local rotations interleaved with nearest-neighbor controlled-phase interactions (see Fig. 3a). 
Second, we consider a trapped-ion ansatz (TIA) composed of the native gates for ion-based quantum devices (Fig. 3b). The probe state is prepared by a unitary composed of interleaving local rotations around the X- and Y-axes, with Mølmer–Sørensen interactions which generate entanglement between adjacent qubits in the chain. In both the HEA and TIA cases, the “brick-work” circuit layer is repeated $d$ times, and has open boundary conditions (i.e., the 1st and $n$-th qubits do not interact).
The final probe state ansatz we consider is a class of graph states amenable to quantum sensing \cite{shettellGraphStatesResource2020a}. Graph states are a class of entangled quantum states especially relevant for photonic quantum technologies \cite{thomasEfficientGenerationEntangled2022a}, as well as other architectures \cite{lanyonMeasurementBasedQuantumComputation2013}. Certain families of graph states, including cluster and star states, have low QFI; however, Shettel \textit{et al.} show that so-called bundled graph states have QFI which scales with the HL multiplied by a constant factor, $1/k$ \cite{shettellGraphStatesResource2020a}. In Fig. 3c, we demonstrate the VQS on a probe state that is locally equivalent to a $k=2$ bundled graph state (corresponding to a complete bipartite graph). Fig. 3 demonstrates the quantum and classical Fisher Information of the sensing circuits during optimization, as well as the performance (i.e., bias and variance) of the neural network estimator after training. 

In the Fisher Information optimization curves (Fig. 3, second row), we use the CFI as the objective function. We observe that the CFI is a more `rugged' objective landscape compared to the QFI -- the spurious decreases during optimization in the CFI are a result of gradient ascent update directions for $\bm{\mu}$ which cause the projective measurement bases to non-optimally capture the large QFI of the parameterized probe state into classical information. Careful selection of the hyperparameters, e.g., the learning rate, is necessary for realizing optimization trajectories that do not get stuck in sub-optimal parameter regimes and converge in an efficient number of steps. We note that using the Fisher Information as an objective function demonstrates barren plateaus, a common and well-known challenge of parameterized quantum circuits and variational quantum algorithms.

The performance of the neural network estimator for all probe state ansätze (Fig. 3, bottom rows) tends towards being unbiased for longer sequence lengths. The variance of the trained VQS protocol reaches below the SQL in the finite data regime and partially saturating the HL. However, at longer sequence lengths tested ($m\sim1000$), the variance begins to increase and no longer reaches the HL. This is due to the combination of discretizing the posterior distribution and the Bayesian update, which leads to spurious noise in $p(\phi_j | \bm{s}^{(m)})$ that is amplified as $m$ increases.

\subsection{Scaling of VQS protocols \label{subsec:scaling}}
After demonstrating the end-to-end variational optimization of the quantum sensing protocols, we turn our attention to exploring the scaling performance of the framework. In Fig. 4a we show the scaling in the number of qubits, from $n=2$ to $n=10$, demonstrating for each that the method can realize estimation precision approaching the HL in noise-free environments. We also observe that 
as $n$ increases, fewer measurement samples are required for the estimator to plateau towards being unbiased, as each individual measurement sample contains more information and correlations between qubits that are useful for the estimation. The exponential scaling of the Hilbert space limits the size of quantum sensors that can be simulated classically; however, variational and optimal quantum control algorithms for NISQ devices have been proposed and demonstrated for driving system dynamics towards states with large Fisher Information, \textcolor{\markupcolor}{and advances in this domain can be readily adopted for sensing protocols}. 
\textcolor{\markupcolor}{Due to noisy circuits and limited shot rates, the capability to sample high-quality training datasets may be challenging for some current qubit architectures. However, as quantum sensing devices improve and scale, we expect such neural network approaches will be vital for sensing protocols. Such models, in general, provide an important tool to counter bias from limited training sets, especially when the state space dimension is much larger than the size of the available dataset.}


In Fig. 4b, we characterize the number of measurement samples per phase needed to robustly train the neural network estimator. As outlines in Section \ref{subsec:device}, the total number of measurement samples in the training dataset is $|\mathcal{D}| = |\mathcal{D}_{\phi}| \times n_{\phi}$. For small training dataset sizes (light green lines), the estimator cannot be robustly trained, leading to the bias diverging away from zero and the variance collapsing to zero (due to the discretized posterior distribution collapsing towards a Kronecker delta function). We observe that training a modestly-sized network on datasets of size approximately $|\mathcal{D}_{\phi}| \sim 1000$ and $n_{\phi} \sim 100$ resulted in robust and high-performance estimation. However, this requirement is highly dependent on the number of qubits, sensing dynamics, and quality of the training data. 

\subsection{Comparison to GHZ protocol \label{subsec:ghz}}
The GHZ protocol is a well-studied, paradigmatic metrology protocol that is optimal in noise-free environments. Here, we compare the effects of noisy probe state preparation on both a GHZ and VQS protocol. We use analogous noise models affecting the state preparation for the two protocols: following every two-qubit gate in the state preparation circuit, the control and target qubits are acted on by independent dephasing channels, each with coefficient $\gamma\in[0, 1]$ \textcolor{\markupcolor}{(see Methods, Section \ref{sec:methods:noise})}. For the GHZ state preparation, the circuit contains $n-1$ two-qubit gates, while the VQS using the HEA consists of $(n-1)d$ two-qubit interactions. For each value of $\gamma$, we optimize the circuit in the face of the noisy state preparation and sample simulated training and testing datasets from the optimized device. We sample a testing dataset of the GHZ and VQS circuits of the same size and values of $\phi$. For the estimation based on GHZ data, we use a maximum likelihood estimator (where the likelihood is under the assumption of noise-free evolution, see Methods); for the VQS we use a trainable neural network ansatz as before. \textcolor{\markupcolor}{We then directly benchmark and compare the bias and variance of the two sensing protocols.} In Fig. 4, we see that for low noise strengths, the GHZ and VQS have comparable Fisher Information, saturating to the optimal value of $n^2$ \textcolor{\markupcolor}{(fixed to $n=4$)}. As the noise increases, the CFI of the GHZ probe decreases rapidly to $0$ at large $\gamma$, such that the measurement outcomes are completely random and carry no information about the value of $\phi$. The VQS can, however, mitigate the effects of the noise by tuning $\bm{\theta}$ and $\bm{\mu}$ to remain above the SQL on the Fisher Information for larger amounts of environmental noise; in-line with recent demonstrations that parameterized circuits are able to mitigate the entropic effects of the environmental noise in shallow circuits \cite{duschenesCharacterizationOverparameterizationSimulation2024}. 
\textcolor{\markupcolor}{
Analyzing the final estimator performance (Fig. 4c, middle and bottom), the GHZ maximum likelihood estimator quickly becomes biased as the dephasing noise increases.} The neural network estimator, on the other hand, compensates for the noise in its estimation as it learns the effect of the noise channels from the sampled data. As such, the neural network estimator remains unbiased as the noise increases, and instead trades-off for increased variance.

\section{Discussion}
In this work, we introduced a variational framework for quantum sensing protocols, optimizing both the quantum evolution and the classical estimator. We model the protocol stages corresponding to probe state preparation, interaction, and measurement with parameterized quantum circuits, optimizing the Fisher Information towards regimes with maximal achievable precision. Subsequently, we use a neural network as a variational estimator ansatz, which is trained on single labeled measurement outcomes and used in a Bayesian context for performing estimation on arbitrarily long sequences of measurements. 

The Fisher Information is a natural choice for the objective function, as it is a straight-forward and interpretable way of quantifying the utility of a probe state for metrology. However, other important protocol considerations are not encapsulated by the Fisher Information, including, for example, the dynamic range of the sensor -- i.e., the effective range over which the unknown parameter can be estimated.
As such, continued investigations of new objective functions \cite{marciniakOptimalMetrologyProgrammable2022} and hybrid control algorithms \cite{youssryExperimentalGrayboxQuantum2024} may suggest novel approaches towards designing more useful and robust sensing devices.
The end-to-end parameterization and differentiable programming of VQS protocols open promising avenues for using higher-order derivative quantities and probability distribution moments for improved objective functions \cite{maclellanInverseDesignPhotonica}. In addition, the variational nature of such methods may enable new approaches for sensing based on adaptive feedback and active learning concepts \cite{langeAdaptiveQuantumState2023, belliardoModelawareReinforcementLearning2024}.

In parallel, as both variational algorithms and machine learning capabilities advance, with model architectures and compute infrastructures rapidly evolving, finding synergistic overlap with challenges in quantum sensing may prove to be a fruitful direction for development.
In particular, future work should explore alternative neural network architectures and training methods towards variational quantum sensing protocols. Architectures that are implicitly designed for sequential data (e.g., recurrent neural networks, transformers \cite{vaswaniAttentionAllYou2017, melkoLanguageModelsQuantum2024}), as well as approaches developed for approximating complex, high-dimensional probability distributions (e.g., normalizing flows \cite{rezendeVariationalInferenceNormalizing2016a}) are likely well-suited towards quantum sensing protocols.

The insights provided by VQS frameworks can be used directly to identify improved operational settings for realistic, constrained qubit architectures, especially relevant for near-term programmable quantum devices \cite{marciniakOptimalMetrologyProgrammable2022,madsenQuantumComputationalAdvantage2022}. 
Further, future work should investigate how such frameworks can be utilized for the automatic discovery of actionable and human-interpretable \cite{krennConceptualUnderstandingEfficient2021a} mitigation strategies to realize and preserve quantum advantage for the next generation of sensing technologies.


\section{Methods} 

\subsection{Circuit simulations and training}
Quantum circuit simulations were performed with differentiable tensor-network methods, using the TensorCircuit and JAX frameworks \cite{zhangTensorCircuitQuantumSoftware2023a, jax2018github}. All circuits evolve a pure $n$-qubit initial state $\ket{ \rho_0 } = \vert 0 \rangle^{ \otimes n} $ through the parameterized state preparation, $U(\bm{\theta})$. The state preparation is unitary, except for in cases when noise is explicitly considered in the analysis of Fig. 4c. 
The probe state then transforms under local rotations parameterized by the parameter-of-interest, $\phi$. The local quantum channel which imprints $\phi$ onto the state is $K(\phi) = \otimes_{i=1}^n R_z(\phi)$. Finally the perturbed probe state is measured by parameterized projective measurements in rotated bases, implemented as a layer of local rotations followed by projecting each qubit into the computational basis (i.e., eigenvectors of the Z operator). Optimization of the circuit parameters, $\bm{\theta}$ and $\bm{\mu}$, is performed using the ADAM algorithm and the Fisher Information as the objective function (unless otherwise noted, we use the CFI as the loss function, but the QFI can also be used). The HEA, TIA, and bundled graph state circuits (including both $\bm{\theta}$ and $\bm{\mu}$) have $2nd + 4n$, $(3n -1)d + 3n$, and $4n$ total variational parameters, respectively. 

\subsection{Neural network training}
In this work, the network architecture used as the estimator ansätze are fully-connected, feed-forward neural networks. The networks have dimension $(n, h_1, \dots, h_{n_\text{layer}}, w)$, where $n$ is the number of qubits, $h_i$ is the number of nodes in the $i$-th hidden layer, $n_\text{layer}$ is the number of hidden layers, and $w$ is the number of output nodes (i.e., the number of phase discretization points). Each neuron passes information through a nonlinear activation function, here the ReLU function, defined as $\text{max}(x, 0)$. The output layer is normalized to a valid probability distribution using the $\text{softmax}$ function. The corresponding phase label, $\phi_j$, for a measurement sample is encoded as a one-hot encoded vector, i.e., a Kronecker delta function, $\delta_{ij}$. We use the cross-entropy function between the label one-hot vector and the output layer as the loss function, with an added L2 regularization term to improve training and estimator performance. The ADAM optimizer is used for stochastic gradient descent. Neural network estimators were implemented and trained using the JAX framework. 

\newcommand{\parity}[1]{\mathrm{P}\left(#1\right)}
\subsection{Comparison with GHZ protocol}
\label{sec:methods:noise}
Here, the probe state is the highly entangled GHZ state, $\vert \text{GHZ} \rangle = \tfrac{1}{\sqrt{2}} \left( \vert 00\dots 0 \rangle + \vert 11\dots 1 \rangle \right)$, which interacts with the parameter of interest through a local rotation around the Z-axis of the Bloch sphere, $R_z(\phi)$. The GHZ state is prepared via the quantum circuit, 
\[
    \ket{\text{GHZ}} 
    = \left( \prod_{i=2}^n \text{CNOT}_{1,i} \right) H_1 \ket{0}^{\otimes n}.
    \numberthis 
\]
A local dephasing channel is applied on both the control and target qubits for every CNOT, described by the Kraus operators,
\[
    A_0 = \begin{bmatrix} 1 & 0 \\ 0 & \sqrt{1-\gamma} \end{bmatrix}, 
    \quad
    A_1 = \begin{bmatrix} 0 & 0 \\ 0 & \sqrt{\gamma} \end{bmatrix}
    ,
\]
and the state evolves as,
\(
    \rho \rightarrow A_0 \rho A_0^\dagger + A_1 \rho A_1^\dagger
\).
\textcolor{\markupcolor}{We study this prototypical noise model as two-qubit interactions are the dominant source of noise in practically all quantum information processing platforms \cite{doi:10.1126/science.abb2823}.}
A parity measurement of the state in the X basis is performed, $\parity{s_i}$, i.e. calculating the sum modulo-two of the sampled bitstring, $s_i$, such that, 
\[
    \parity{s_i} = 
    \begin{cases}
        0 \quad \text{if $s_i$ is even parity} \\
        1 \quad \text{if $s_i$ is odd parity} 
    \end{cases}
    . \numberthis
\]
The likelihood distribution of measuring an even parity measurement for these sensing dynamics, in the noise-free case, is, 
\begin{align*}
    &p\left(\parity{s_i} = 0 | \phi\right) = \cos^2 \left( \frac{ n \phi}{2} \right)\\
    &p\left(\parity{s_i} = 1 | \phi\right) = \sin^2 \left( \frac{ n \phi}{2} \right).
    \numberthis
\end{align*}
Using a maximum likelihood estimator, i.e., maximum \textit{a posteriori} estimation with a uniform prior distribution, our estimator for a sequence of $m$ measurement samples is,
\begin{align*}
   	p( \bar{\phi} | \bm{s}^{(m)})  \numberthis
    &\propto \prod_{i=k}^m p( \bar{\phi} | s_i) \\
    &= \prod_{k=1}^m \parity{s_i} \cos^2 \left(\frac{n\phi}{2} \right)  + (1 - \parity{s_i}) \sin^2 \left( \frac{n \phi}{2} \right),
\end{align*}
\textcolor{\markupcolor}{with a variance which scales as $\Delta^2\bar{\phi} \propto n^{-2}$ and, as it is derived using the noise-free likelihood, is not dependent on $\gamma$.}
To compare the VQS and GHZ protocols, both are defined as quantum circuits, and projective measurement samples are collected from both over a range of phase values $\phi_k$. Training data is collected for the neural network estimator, and equivalent testing data sets are sampled for both the GHZ and VQS protocols.

\section{Data Availability}
All data related to this work and its results are available from the corresponding author on reasonable request. 

\section{Code Availability}
The code supporting this work is released under the Apache 2.0 license at 
\href{https://github.com/benjimaclellan/queso}{https://github.com/benjimaclellan/queso}.


\section{Acknowledgement}
We thank Matthew Duschenes, Yi Hong Teoh, Sebastian Wetzel, Logan Cooke, Johannes Jakob Meyer, and Augusto Smerzi for critically helpful discussions.
This work was supported by NSERC and the Perimeter Institute for Theoretical Physics.
B.M. acknowledges support from NSERC through the Vanier CGS program.
R.G.M. acknowledges support from NSERC through the Discovery Grant program.
The calculations in this work were enabled in part by support provided by SHARCNET and Compute Canada. 
Research at Perimeter Institute is supported in part by the Government of Canada through the Department of Innovation, Science and Economic Development Canada and by the Province of Ontario through the Ministry of Economic Development, Job Creation and Trade.

{
\color{\markupcolor}
\section{Author Contributions}
B.M. and P.R. conceived the idea. B.M. developed the circuit optimization and machine learning algorithms, and performed the numerical analysis. P.R., S.C., and R.G.M. provided supervision and guidance during the project. All authors contributed to discussions and the writing and editing of the manuscript.

\section{Competing Interests}
The authors declare no competing interests.
}

\bibliography{bib}

\clearpage

\end{document}